\documentclass{article}
\usepackage{iclr2027_conference,times}
\iclrfinalcopy

\usepackage{amsmath,amsfonts,bm}

\def\eqref#1{equation~\ref{#1}}

\def\1{\bm{1}}

\DeclareMathAlphabet{\mathsfit}{\encodingdefault}{\sfdefault}{m}{sl}
\SetMathAlphabet{\mathsfit}{bold}{\encodingdefault}{\sfdefault}{bx}{n}

\newcommand{\benchmark}{\textsc{SimuVerity}}

\newcommand{\simulink}{\textsc{Simulink}}
\newcommand{\slx}{\texttt{.slx}}

\usepackage{graphicx}
\usepackage{booktabs}
\usepackage{multirow}
\usepackage{array}
\usepackage{longtable}
\usepackage{needspace}
\usepackage{etoolbox}
\newlength{\svtablecaptionskip}
\AddToHook{env/table/begin}{\setlength{\belowcaptionskip}{\svtablecaptionskip}}
\newcolumntype{P}[1]{>{\raggedright\arraybackslash}p{#1}}
\newcolumntype{B}[1]{>{\bfseries\raggedright\arraybackslash}p{#1}}
\newcommand{\svtablehead}[1]{\textbf{\boldmath #1}}
\newcommand{\svtablestyle}{\renewcommand{\arraystretch}{1.08}}
\makeatletter
\let\sv@LT@makecaption\LT@makecaption
\patchcmd{\sv@LT@makecaption}{\vskip\baselineskip}{\vskip\svtablecaptionskip}{}%
  {\PackageError{simuverity}{Longtable caption spacing patch failed}{Check the longtable caption definition.}}
\long\def\LT@makecaption#1#2#3{%
  \sv@LT@makecaption{#1}{\normalsize #2}{\normalsize #3}}
\makeatother
\usepackage{float}
\usepackage{xcolor}
\usepackage{hyperref}
\usepackage{url}

\graphicspath{{figures/}}

\title{SimuVerity: Benchmarking Agents for\\Engineering-Grade Simulink\\Model Generation}

\author{%
Ruiqi Zhang$^{*}$ \quad Jiahao Wang$^{*}$ \quad Mingxuan Li \quad Haichen Luo \\
\bfseries Chaoting Wang \quad Guoyu Mou \quad Keyu Lai \quad Hanchao Lv \\
\bfseries Jiaxu Wang \quad Yibo Zheng \quad Aijun Yang$^{\dagger}$ \quad Xiaohua Wang$^{\dagger}$ \\
\normalfont Xi'an Jiaotong University \\
\normalfont Code and data: \url{https://github.com/SimuVerity/SimuVerity}%
}
\makeatletter
\def\@maketitle{\vbox{\hsize\textwidth\centering
{\LARGE\sc \@title\par}
\begin{tabular}[t]{c}\bf\rule{\z@}{24pt}\@author\end{tabular}\par
\vskip 0.3in minus 0.1in}}
\newcommand{\svauthornotes}{\g@addto@macro\@thanks{\begin{NoHyper}\footnotetext[0]{$^{*}$Equal contribution. \quad $^{\dagger}$Corresponding authors.}\end{NoHyper}}}
\makeatother

\begin{document}

\svauthornotes
\maketitle
\lhead{Preprint}

\begin{abstract}
Existing \simulink{} benchmarks mainly evaluate whether generated models compile, execute, or resemble a reference model. These criteria do not establish whether a model satisfies its engineering requirements. We introduce \benchmark{}, a benchmark of 101 text-to-executable \simulink{} model-generation tasks across ten engineering domains. For each task, executable-system profiles ground the engineering specification and four families of native simulation scenarios. A hierarchical evaluator first checks artifact delivery, native executability, and engineering qualification, then scores qualified models across six dimensions covering accuracy, output quality, mechanistic fidelity, control and causal integrity, operating-domain robustness, and dynamic response. We evaluate six agent systems with \benchmark{}. The best system achieves an overall score of only 42.86. The results show that structural similarity is a poor proxy for engineering performance: capability bottlenecks arise both in producing qualified implementations and in satisfying multidimensional requirements after qualification. Meanwhile, some high-scoring models still exhibit severe visual-layout disorder. \benchmark{} provides a systematic basis for assessing agents' engineering capabilities and diagnosing failures in executable \simulink{} model generation.

\end{abstract}

\section{Introduction}
\label{sec:introduction}

\begingroup
\setlength{\intextsep}{4pt}
\begin{figure}[H]
  \centering
  \includegraphics[width=\linewidth]{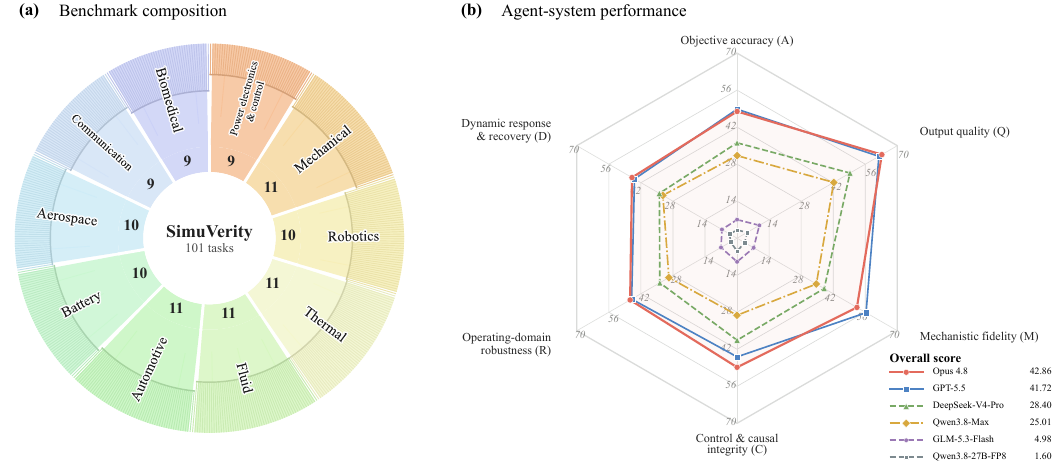}
  \caption{\textbf{Task composition and agent performance in \benchmark{}.} (a) Task distribution across ten engineering domains; outer-ring line lengths indicate mean domain scores. (b) Six-dimensional performance and overall scores of six agents.}
  \label{fig:overview}
\end{figure}
\endgroup

\simulink{} is a powerful simulation and Model-Based Design environment for system modeling, dynamic analysis, control design, and verification, with broad use across industrial engineering domains \citep{shrestha2022slnet,mathworks2026simulink}. Recent LLM-based agents can translate natural-language requirements into executable \slx{} models, run simulations, analyze results, and iteratively refine their implementations \citep{abdalla2026aswe,liang2026simuagent}. Therefore, as engineering practice demands high system accuracy, performance, and robustness, an important question is how to benchmark whether these agents produce models that truly satisfy engineering requirements rather than merely execute.

However, existing benchmarks have not centered the core engineering question of whether a generated model satisfies its requirements across prescribed operating conditions. They primarily rely on two types of criteria: (i) artifact delivery and technical execution, and (ii) structural and functional similarity to a reference system \citep{shrestha2021slgpt,ren2025simugen,liang2026simuagent}. But the former only verifies that a model exists and runs, while the latter treats one reference implementation as the standard despite valid structural alternatives and possible errors in reference-aligned models. Some automotive studies improve on these criteria with requirement-level tests, but remain limited to software logic and control functions \citep{abdalla2026aswe}. A cross-domain benchmark must therefore evaluate candidates against their engineering requirements through task-specific native simulation scenarios and multidimensional performance criteria over the prescribed operating domain.

To address this gap, we introduce \benchmark{}, an engineering-grade native-simulation benchmark with 101 tasks across ten engineering domains. For each task, domain experts build an executable-system profile from native runs of a selected, reconstructed, or independently implemented reference system. The profile captures system boundaries, interfaces, engineering relations, operating ranges, and dynamics, grounding both the natural-language specification and four families of task-specific simulation scenarios. Evaluation first checks artifact delivery, technical executability, and engineering qualification, then scores qualified candidates across A/Q/M/C/R/D. This design supports multidimensional diagnosis, valid structural diversity, and detection of reference-aligned models with poor engineering performance.

We evaluate six agent systems with \benchmark{}. The highest-scoring system reaches only 42.86 overall, indicating limited capability on multi-domain \simulink{} engineering tasks. Its artifact-delivery, executability, and engineering-qualification rates are 94.72\%, 86.47\%, and 76.90\%, respectively. Thus, some delivered and executable candidates do not form qualified engineering implementations, and shallow success cannot replace multidimensional evaluation. We further analyze potential deficiencies in the generated models. Experts classify eight of the 31 candidates from the highest-scoring agent system with engineering scores of at least 70 as severely disordered, showing that agents often overlook readability despite strong engineering performance. Current agents therefore lack effective cross-domain \simulink{} generation capability and require a closed loop of specification-driven modeling, native simulation validation, and multi-condition feedback correction.

In conclusion, our main contributions are:
\begin{enumerate}
  \renewcommand{\labelenumi}{(\Roman{enumi})}
  \item \textbf{Benchmark curation.} We introduce \benchmark{}, a benchmark of 101 text-to-executable \simulink{} model-generation tasks across ten engineering domains. Each task includes a reconstructed or independently built reference system, an executable-system profile, and a profile-grounded engineering prompt.
  \item \textbf{Evaluation framework.} We design a hierarchical multidimensional performance evaluation framework comprising four families of task-specific native simulation scenarios and six performance dimensions. It enables hierarchical assessment and multidimensional diagnosis of candidate engineering performance.
  \item \textbf{Empirical findings.} We evaluate six agent systems and analyze the performance and limitations of their generated models from multiple engineering perspectives, laying an empirical foundation for future improvements in \simulink{} model-generation performance.
\end{enumerate}

\section{Related Work}
\label{sec:related-work}

\noindent\textbf{Benchmarks for executable engineering artifacts.}
Domain-specific benchmarks increasingly evaluate complete executable engineering artifacts through native execution and requirement-level evidence, moving beyond artifact formation and basic validity checks \citep{chen2025scienceagentbench,guo2025engdesign,cui2026hwebench,qin2026structureclaw,dong2026muse,singh2026cadengbench,chen2021humaneval,liu2023verilogeval,lu2024rtllm}. This progression raises the corresponding question for executable dynamical systems: how should native execution evidence support a hierarchical, multidimensional assessment of engineering compliance?

\noindent\textbf{Agentic \simulink{} model generation and evaluation.}
Existing \simulink{} studies target toolchain fault discovery (SLGPT), block-diagram reconstruction with basic execution checks (SimuGen), or automotive control validation with boundary and time-series tests (ASWE-Bench) \citep{shrestha2021slgpt,ren2025simugen,abdalla2026aswe,kevian2024controlbench,guo2024controlagent}. SimuBench extends text-to-model generation across multiple domains, but relies mainly on artifact completeness, basic executability, and structural or functional similarity to a reference model \citep{liang2026simuagent}. Such criteria can miss structurally distinct valid implementations and engineering failures in reference-aligned candidates. Native MATLAB/\simulink{} access for agents further broadens the space of valid implementations \citep{mathworks2026matlabmcpserver,mathworks2026simulinkagentictoolkit}. \benchmark{} addresses these gaps with executable-system-profile-grounded simulation scenarios and hierarchical multidimensional evaluation. Detailed comparisons appear in Appendix~\ref{app:closest-work}.

\raggedbottom

\begin{figure}[!t]
  \centering
  \includegraphics[width=\linewidth]{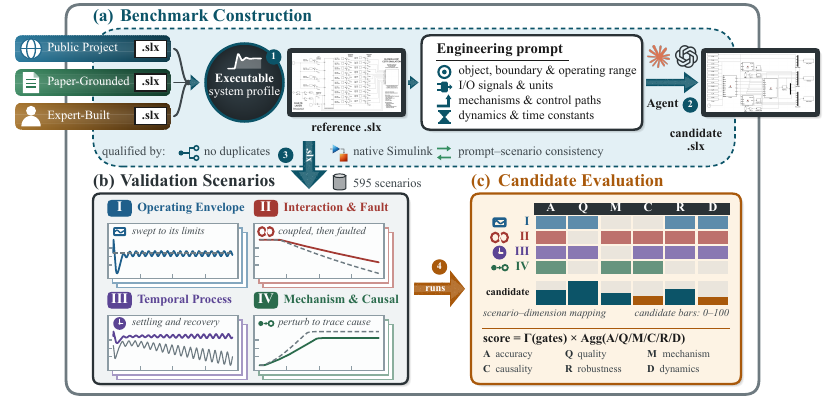}
  \caption{\textbf{The \benchmark{} pipeline.} Benchmark construction, native simulation testing, and six-dimensional performance evaluation.}
  \label{fig:pipeline}
\end{figure}

\section{\benchmark{}}
\label{sec:benchmark}

\subsection{Task Definition and Cross-Domain Coverage}
\label{sec:scope-taxonomy}

\paragraph{Task Definition.}

We formulate text-to-executable \simulink{} model generation as a build-to-specification task. Given a natural-language engineering specification defining the system, interfaces, operating conditions, functional objectives, mechanisms, and dynamics, a system must deliver a complete native \slx{} model. Block choices, diagram layout, and subsystem hierarchy remain unrestricted, allowing diverse valid implementations.

\paragraph{Cross-Domain Coverage.}
\benchmark{} comprises 101 tasks spanning ten engineering domains: power electronics and control, mechanical systems, robotics, thermal systems, fluid systems, automotive systems, battery systems, aerospace systems, communication systems, and biomedical systems. Each domain contributes 9--11 tasks, yielding a near-balanced cross-domain composition (Table~\ref{tab:domain-composition}).

\begin{table}[!t]
  \caption{\textbf{Task distribution across the ten engineering domains of \benchmark{}.}}
  \label{tab:domain-composition}
  \centering
  \small
  \begingroup
  \setlength{\tabcolsep}{6pt}
  \svtablestyle
  \begin{tabular*}{\linewidth}{@{\extracolsep{\fill}}lr@{\hspace{2.5em}}lr@{}}
    \toprule
    \multicolumn{4}{c}{\svtablehead{Engineering domains} (101 tasks)} \\
    \midrule
    Power electronics \& control & 9 & Automotive systems & 11 \\
    Mechanical systems & 11 & Battery systems & 10 \\
    Robotics & 10 & Aerospace systems & 10 \\
    Thermal systems & 11 & Communication systems & 9 \\
    Fluid systems & 11 & Biomedical systems & 9 \\
    \bottomrule
  \end{tabular*}
  \endgroup
\end{table}

\subsection{Data Curation}
\label{sec:curation}

\paragraph{Reference-System Curation and Reconstruction.}
We collect candidate reference systems from public examples and engineering projects, research artifacts accompanying papers, and expert-built models. We retain systems that can be executed in \simulink{}, represent meaningful engineering behavior, and broaden the benchmark's coverage. Each selected system is then reconstructed into a reproducible evaluation target by resolving dependencies and compatibility issues, exposing the signals needed for control and observation, and connecting it to repeatable operating conditions and a task-specific evaluator.

\paragraph{Executable-System Profiling and Prompt Design.}
Domain experts inspect and execute each reference system under a unified protocol to construct an executable-system profile grounded in observable model facts and reproducible simulation evidence. The profile captures the system boundary, interfaces, mechanisms, control relationships, operating conditions, and dynamic behavior. Based on these profiles, human experts design a unified prompt framework covering the modeled system, deliverable, input--output interfaces, operating range, functional objectives, required mechanisms, control relationships, and dynamic behavior. They use this framework to write a prompt for each task. Each prompt is then checked against its executable-system profile for engineering consistency and native \simulink{} realizability.

\paragraph{Quality Control.}
We track source lineage and apply same-source deduplication and near-duplicate screening. Task-specific test conditions, scenario combinations, thresholds, and scoring logic are withheld from the evaluated agents during model generation to reduce retrieval- and memorization-based shortcuts. A task enters the formal benchmark only after its reference system completes native loading, diagram updating, compilation, and simulation under a unified toolchain. We then verify that the observed system behavior supports the executable-system profile, the public specification accurately reflects that profile, and the independent evaluator tests the specified requirements through the correct input--output interfaces. The final benchmark contains 101 tasks.

\subsection{Task-Specific Simulation Test Design}

\label{sec:native-tests}

This section translates each task's engineering requirements into executable simulation tests that produce the evidence needed for evaluation.

For each task, domain experts translate the requirements in the public engineering specification and executable-system profile into an executable test suite. Each scenario defines the conditions to control, the external and internal signals to observe, the engineering quantities to compute, and the criteria used to interpret them.

\benchmark{} organizes these tests into four complementary families:
\begin{enumerate}
    \item[(i)] \textbf{Operating-envelope scenarios} test nominal operation, boundary conditions, parameter and load variations, and external disturbances.
    \item[(ii)] \textbf{Interaction and fault scenarios} test joint inputs, compound operating conditions, fault injection, and transitions between normal and faulty states.
    \item[(iii)] \textbf{Temporal-process scenarios} test startup, transients, settling, event order, fault onset and clearance, and recovery.
    \item[(iv)] \textbf{Mechanism and causal-check scenarios} test whether required physical paths, functional mechanisms, feedback loops, control relations, and signal provenance are active and correct.
\end{enumerate}

For each applicable scenario, the candidate model is loaded, updated, compiled, and simulated in the native MATLAB/\simulink{} runtime. The resulting execution status, trajectories, engineering quantities, and mechanism checks constitute the evidence used by the evaluation framework in Section~\ref{sec:evaluator}.

\begin{figure}[!t]
  \centering
  \includegraphics[width=\linewidth]{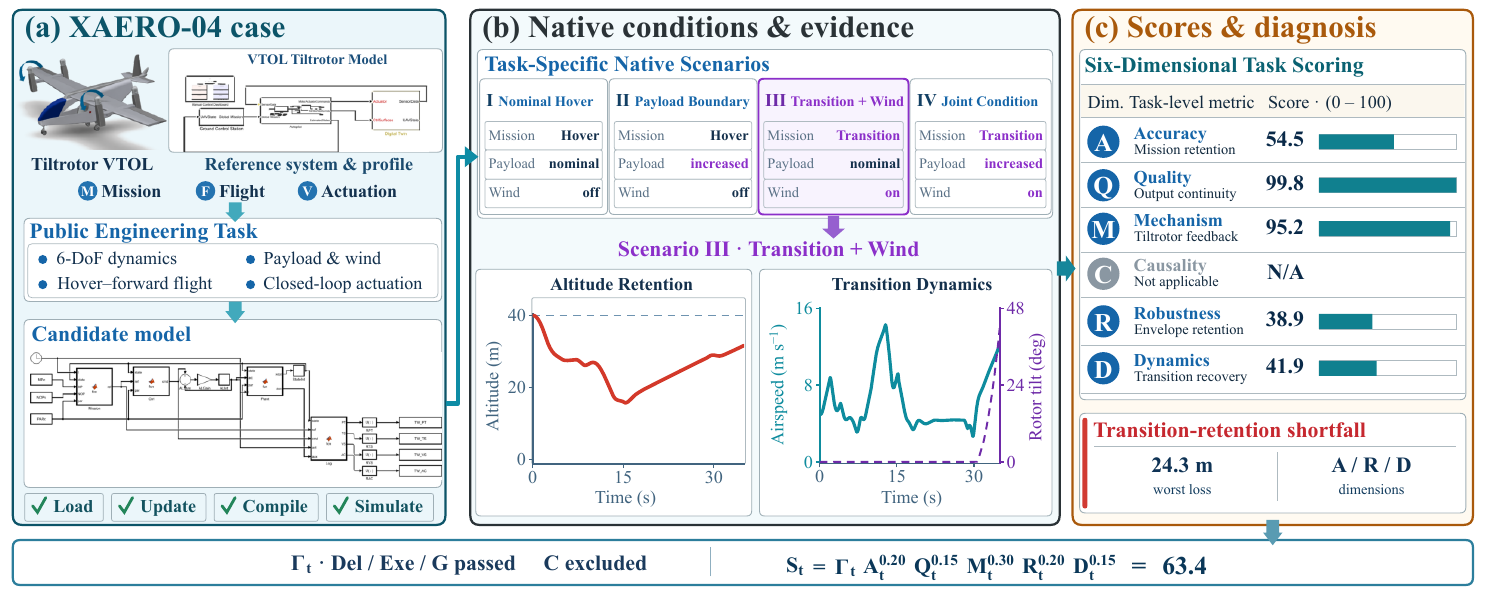}
  \caption{\textbf{Evaluation of the closed-loop tiltrotor VTOL model.} Native simulation evidence, six-dimensional scores, and failure diagnosis for the candidate model. Four of six scenarios are shown.}
  \label{fig:task-walkthrough}
\end{figure}

\subsection{Hierarchical Multidimensional Performance Evaluation Framework}
\label{sec:evaluator}

\benchmark{} first applies three prerequisite gates and then scores qualified candidates across six engineering dimensions. For candidate model $x$ on task $t$, the complete evaluation is
\[
\mathcal{V}_t(x)
=
\left(r_t^{\mathrm{del}},r_t^{\mathrm{exe}},r_t^G,\mathbf{v}_t\right).
\]

\paragraph{Prerequisite Gates.}
$r_t^{\mathrm{del}}$ checks whether the required \texttt{.slx} artifact is delivered, and $r_t^{\mathrm{exe}}$ checks whether it completes native loading, diagram updating, compilation, and simulation. The engineering-qualification gate $r_t^G$ evaluates only candidates that have been delivered and can execute natively, and returns pass or fail. Its checks are specified separately for each task by human experts based on the task specification and executable-system profile: the required functional roles must be present, the required signal paths must be connected, and the candidate must produce valid evaluation evidence in the formal scenarios according to the prescribed output conventions. Candidates that pass the G gate proceed to the subsequent six-dimensional performance scoring.

\paragraph{Multidimensional Performance.}
Candidates passing all three gates receive the performance vector
\[
\mathbf{v}_t
=
\left(v_t^A,v_t^Q,v_t^M,v_t^C,v_t^R,v_t^D\right),
\qquad \mathcal{P}=\{A,Q,M,C,R,D\}.
\]
The six dimensions measure \textbf{A}, objective accuracy; \textbf{Q}, output quality; \textbf{M}, mechanistic fidelity; \textbf{C}, control and causal integrity; \textbf{R}, operating-domain robustness; and \textbf{D}, dynamic response and recovery.

\paragraph{Evidence Mapping and Scoring.}
Let $\mathcal{S}_t$ be the scenario set and $\mathcal{K}_t$ the observed engineering quantities for task $t$. The task-specific relation
\[
\mathcal{R}_t\subseteq\mathcal{S}_t\times\mathcal{K}_t\times\mathcal{P}
\]
maps each quantity $k$ observed under scenario $s$ to a performance dimension $d$. This mapping is many-to-many: one scenario may support several dimensions, and one dimension may combine evidence from several scenarios.

\begin{table}[!t]
  \caption{\textbf{Overall scores and prerequisite pass rates.} Agent results are three-run means; Reference reports the mean score of 101 task-specific reference systems.}
  \label{tab:overall-results}
  \centering
  \small
  \setlength{\tabcolsep}{5pt}
  \svtablestyle
  \begin{tabular*}{\linewidth}{@{\extracolsep{\fill}}lrrrr@{}}
    \toprule
    \svtablehead{Agent system} & \svtablehead{Overall} & \svtablehead{Delivered} & \svtablehead{Executable} & \svtablehead{G pass} \\
    \midrule
    Opus 4.8 + Claude Code        & 42.86 & 94.72\% & 86.47\% & 76.90\% \\
    GPT-5.5 + Codex               & 41.72 & 98.68\% & 94.72\% & 75.58\% \\
    DeepSeek-V4-Pro + Claude Code & 28.40 & 97.03\% & 88.12\% & 61.39\% \\
    Qwen3.8-Max + Claude Code     & 25.01 & 89.77\% & 80.86\% & 53.80\% \\
    GLM-5.3-Flash + Claude Code   & 4.98  & 96.04\% & 48.18\% & 13.20\% \\
    Qwen3.8-27B-FP8 + Claude Code & 1.60  & 30.36\% & 13.53\% & 6.27\% \\
    \midrule
    \svtablehead{Reference}                     & 96.01 & 100.00\% & 100.00\% & 100.00\% \\
    \bottomrule
  \end{tabular*}
\end{table}

Each observed quantity $z_{t,s,k}$ is converted to a normalized utility by a task-specific function,
\[
u_{t,s,k}=\phi_{t,s,k}\!\left(z_{t,s,k}\right),
\qquad u_{t,s,k}\in[0,1].
\]
The relevant utilities are then aggregated into each dimension:
\[
v_t^d
=
\operatorname{Agg}_{t,d}
\!\left(\left\{u_{t,s,k}\mid(s,k,d)\in\mathcal{R}_t\right\}\right).
\]
The resulting dimension scores are reported on a 0--100 scale. The final task score is
\[
S_t(x)
=
\Gamma_t\!\left(r_t^{\mathrm{del}},r_t^{\mathrm{exe}},r_t^G\right)
H_t\!\left(\mathbf{v}_t\right),
\qquad
H_t(\mathbf{v}_t)
=
\prod_{d\in\mathcal{P}_t}
\left(v_t^d\right)^{w_{t,d}},
\]
where $\mathcal{P}_t\subseteq\mathcal{P}$ is the applicable dimension set for task $t$, $\sum_{d\in\mathcal{P}_t}w_{t,d}=1$, and $\Gamma_t$ applies the prerequisite gates and $H_t$ aggregates the six-dimensional performance. When required by the task, worst-condition or bottleneck-sensitive aggregation prevents critical failures from being hidden by strong results on easier conditions. The evaluator returns the prerequisite outcomes, six-dimensional performance, overall score, and the scenarios and quantities responsible for each failure.
Figure~\ref{fig:task-walkthrough} illustrates this evaluation chain for the closed-loop tiltrotor VTOL task.

\paragraph{Expert Validation of Evaluator Validity.}
To test whether evaluator scores reflect engineering judgment~\citep{zhu2025agenticbenchbestpractices}, two domain experts independently rated 30 candidate models generated by Opus 4.8 Run~1. The candidates span all ten engineering domains and were rated on a five-point scale without access to evaluator scores. Inter-rater agreement was high, with weighted Cohen's $\kappa=0.90$. The Spearman correlation between evaluator scores and mean expert ratings was $\rho=0.94$, comparable to the inter-expert rank agreement ($\rho=0.87$). Even when restricted to the 18 candidates with nonzero evaluator total scores, the correlation remained $\rho=0.88$. These results indicate close agreement between evaluator rankings and domain-expert judgments. Detailed sampling and rank-correlation results appear in Appendix~\ref{subsec:expert-validation}.

\section{Experiments and Results}
\label{sec:experiments}

\subsection{Experimental Setup}
\label{sec:experimental-setup}

\paragraph{Agent Systems.}
We evaluate six agent systems: Claude Opus 4.8~\citep{anthropic2026opus48} with Claude Code~\citep{anthropic2026claudecode}, GPT-5.5~\citep{openai2026gpt55} with Codex~\citep{openai2026codexcli}, DeepSeek-V4-Pro~\citep{deepseek2026v4pro} with Claude Code, Qwen3.8-Max~\citep{qwen2026qwen38max} with Claude Code, GLM-5.3-Flash~\citep{zai2026glm53flash} with Claude Code, and Qwen3.8-27B-FP8~\citep{qwen2026qwen3827bfp8} with Claude Code. Harness versions, reasoning settings, and context configurations are reported in Appendix~\ref{subsec:implementation-settings}.

\paragraph{Common Task and Execution Conditions.}
All six systems are evaluated on the same 101 tasks across ten engineering domains, using identical public specifications, delivery requirements, task workspaces, and MATLAB/\simulink{} environments. Each run begins in a clean session with an independent working directory and access to the same core MATLAB MCP tools. These tools connect the agent to the native MATLAB/\simulink{} runtime for model construction, inspection, and testing~\citep{mathworks2026matlabmcpserver,mathworks2026simulinkagentictoolkit,yang2024sweagent}.

\paragraph{Execution Environment and External-Resource Isolation.}
The execution environment is isolated from the outside: the example-model directories of the MATLAB installation, the local reference-model directories, and the scenario and scoring-script directories are invisible to the agent.

\paragraph{Evaluation Protocol.}
We impose common stopping conditions and delivery rules on all systems~\citep{kapoor2024aiagentsmatter}. Each task--system pair is run three times from a fresh session and directory, with a separate run identifier. The final \texttt{.slx} artifact from each run is evaluated using the same frozen task-specific scorer. We average the three runs to obtain the task--system score, then average across tasks for system-level and domain-level results; each A/Q/M/C/R/D dimension is averaged over its registered applicable tasks. Non-delivery, technical non-execution, and candidate-caused runtime failures count as failures. Errors attributable to the API, evaluator, or execution environment are classified as infrastructure failures and rerun under a new identifier.

\subsection{Overall Performance}
\label{sec:main-results}

\paragraph{Overall System Performance.}
Table~\ref{tab:overall-results} summarizes the overall score and three prerequisite outcomes for all six agent systems. Claude Opus 4.8, GPT-5.5, DeepSeek-V4-Pro, Qwen3.8-Max, GLM-5.3-Flash, and Qwen3.8-27B-FP8 obtain overall scores of 42.86, 41.72, 28.40, 25.01, 4.98, and 1.60, respectively. Even the highest-scoring agent system remains far from fully satisfying the engineering requirements across the benchmark. Task-level 95\% CIs are given in Table~\ref{tab:appendix-main-reliability}.

\paragraph{Prerequisite Gate Outcomes.}
Across all task--system runs, artifact delivery, native executability, and G-pass rates are 84.43\%, 68.65\%, and 47.85\%, respectively. Thus, delivery and executability alone do not guarantee a qualified engineering implementation.

\paragraph{Multidimensional Performance Profiles.}
Table~\ref{tab:dimension-profiles} reports end-to-end and post-G performance across A/Q/M/C/R/D. End-to-end results capture the full generation pipeline: a run that fails the G gate contributes zero to every applicable dimension. Post-G results isolate the performance of qualified engineering implementations by averaging the G-passing candidates within each run. Averaged equally across the six systems, end-to-end A/Q/M/C/R/D scores are 29.03, 38.46, 31.99, 29.21, 27.69, and 27.82, respectively. Output quality is the highest, and the other five dimensions are lower, with operating-domain robustness (R) and dynamic response and recovery (D) the lowest.

\begin{table}[!t]
  \caption{\textbf{End-to-end and post-G performance on the six engineering dimensions.} Each dimension is averaged over its registered applicable-task set.}
  \label{tab:dimension-profiles}
  \centering
  \small
  \setlength{\tabcolsep}{4.5pt}
  \svtablestyle
  \begin{tabular*}{\linewidth}{@{\extracolsep{\fill}}llrrrrrr@{}}
    \toprule
    \svtablehead{Agent system} & \svtablehead{Scope} & \svtablehead{A} & \svtablehead{Q} & \svtablehead{M} & \svtablehead{C} & \svtablehead{R} & \svtablehead{D} \\
    \midrule
    Opus 4.8 + Claude Code
      & End-to-end & 47.95 & 63.24 & 52.34 & 48.82 & 46.72 & 45.90 \\
      & Post-G     & 62.28 & 82.21 & 65.25 & 58.92 & 60.73 & 58.65 \\
    \addlinespace
    GPT-5.5 + Codex
      & End-to-end & 48.80 & 61.94 & 56.34 & 44.87 & 46.05 & 44.81 \\
      & Post-G     & 64.57 & 81.96 & 72.43 & 53.62 & 60.92 & 59.07 \\
    \addlinespace
    DeepSeek-V4-Pro + Claude Code
      & End-to-end & 36.08 & 49.31 & 38.05 & 38.57 & 33.75 & 33.93 \\
      & Post-G     & 58.78 & 80.32 & 59.32 & 57.59 & 55.07 & 54.50 \\
    \addlinespace
    Qwen3.8-Max + Claude Code
      & End-to-end & 31.24 & 42.20 & 34.62 & 29.15 & 29.77 & 32.41 \\
      & Post-G     & 58.03 & 78.46 & 60.57 & 48.47 & 55.40 & 59.14 \\
    \addlinespace
    GLM-5.3-Flash + Claude Code
      & End-to-end & 7.03 & 9.72 & 7.25 & 8.92 & 7.02 & 6.60 \\
      & Post-G     & 54.23 & 75.01 & 50.62 & 55.33 & 53.45 & 52.14 \\
    \addlinespace
    Qwen3.8-27B-FP8 + Claude Code
      & End-to-end & 3.09 & 4.38 & 3.34 & 4.91 & 2.81 & 3.28 \\
      & Post-G     & 49.51 & 68.50 & 51.81 & 57.25 & 45.14 & 51.00 \\
    \bottomrule
  \end{tabular*}
\end{table}

\begin{table}[!t]
  \caption{\textbf{Structural relation and engineering performance.} Results cover 95 delivered Opus Run~1 candidates; post-G means include only candidates that pass engineering qualification.}
  \label{tab:similarity-performance}
  \centering
  \small
  \setlength{\tabcolsep}{4.5pt}
  \svtablestyle
  \begin{tabular*}{\linewidth}{@{\extracolsep{\fill}}lrrrrrr@{}}
    \toprule
    \svtablehead{Structural relation} & \svtablehead{$n$} & \svtablehead{G pass} & \svtablehead{End-to-end} & \svtablehead{Post-G} & \svtablehead{Top quartile} & \svtablehead{Zero} \\
    \midrule
    Reference-Aligned     & 39 & 87.18\% & 52.13 & 59.80 & 12 & 7 \\
    Partially Similar     & 43 & 83.72\% & 44.05 & 52.61 & 9  & 10 \\
    Structurally Distinct & 13 & 69.23\% & 42.37 & 61.20 & 3  & 4 \\
    \bottomrule
  \end{tabular*}
\end{table}

\subsection{Structural Similarity and Engineering Performance}
\label{sec:evaluator-results}

Section~\ref{sec:evaluator} specifies that candidate performance is adjudicated by the hierarchical multidimensional performance framework rather than scored by similarity to a reference system. We test whether candidate--reference structural similarity can serve as a reliable criterion for engineering performance. Domain experts classify the 95 formally delivered Opus candidates from Run~1 into three mutually exclusive structural relations---\textbf{Reference-Aligned}, \textbf{Partially Similar}, and \textbf{Structurally Distinct}---using block choices and abstraction level, system topology, subsystem decomposition, and signal and feedback routing. The groups contain 39, 43, and 13 candidates, respectively; 56/95 (58.95\%) are Partially Similar or Structurally Distinct.

Table~\ref{tab:similarity-performance} compares the groups. Their end-to-end means are 52.13, 44.05, and 42.37, while their post-G means are 59.80, 52.61, and 61.20 for Reference-Aligned, Partially Similar, and Structurally Distinct models, respectively. Of the 24 top-quartile candidates, 12 are Partially Similar or Structurally Distinct, whereas seven Reference-Aligned candidates receive zero, including two that pass G. Thus, the results show both valid structural diversity and deceptive structural similarity: reference proximity does not reliably order engineering performance.

The representative valid-diversity and deceptive-similarity cases are reported in Appendix~\ref{app:evaluator-controls}. They present two Structurally Distinct candidates that pass G and score 85.10 and 80.07, and two Reference-Aligned candidates that pass G but score zero.

\subsection{Cross-Domain Capability Gaps}
\label{sec:cross-domain}

We analyze Claude Opus~4.8 with Claude Code across ten engineering domains by jointly considering engineering-qualification pass rates and post-G performance. The full domain profiles are reported in Appendix~\ref{app:full-results} (Table~\ref{tab:domain-regimes}). The results reveal three capability-gap patterns: qualification-dominated domains such as power electronics and fluid systems; post-qualification bottlenecks in battery and aerospace systems; and mixed bottlenecks in mechanical and communication systems. These patterns show that progress is needed both to form genuine engineering implementations and to satisfy multidimensional requirements after qualification.

\subsection{Visual Layout Quality of High-Performing \simulink{} Models}
\label{sec:diagram-maintainability}

To examine whether high engineering performance is accompanied by readable diagrams, domain experts inspected high-scoring Run~1 candidates from Claude Opus~4.8 with Claude Code. Among the 31 candidates with engineering scores of at least 70, the experts identified several cases with substantial block overlap, difficult-to-trace signal routes, or poorly organized functional regions; eight were judged severely disordered. We then quantified these cases using block overlap, routing detours, and line-crossing density (defined in Appendix~\ref{app:layout-metrics}), and produced layout-only reorganizations for direct before-and-after comparison, adjusting only block positions and existing line routes while keeping each model's implementation unchanged.

\begin{figure}[!t]
  \centering
  \includegraphics[width=\linewidth]{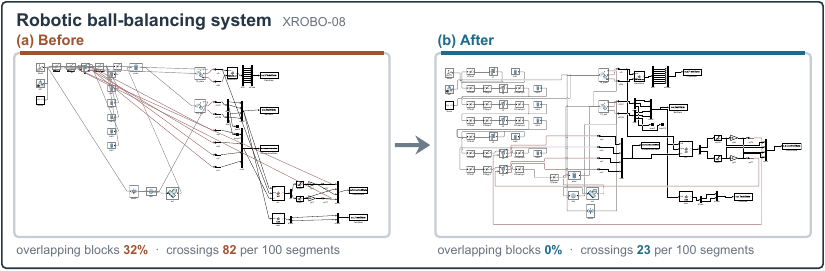}
  \caption{\textbf{Robotic ball-balancing system before and after layout optimization.} Only block positions and existing line routes are adjusted; the model implementation remains unchanged.}
  \label{fig:visual-layout}
\end{figure}

Across the eight cases, reorganization reduced block overlap to zero in every model, reduced line-crossing density by 22.7--85.1\%, and reduced routing-detour ratio in seven cases. Figure~\ref{fig:visual-layout} shows one representative before-and-after example, the robotic ball-balancing system; the complete measurements, intervention protocol, and remaining cases are provided in Appendix~\ref{app:visual-layout}. These results show that high engineering scores do not imply clear diagram organization.

\subsection{Ablations of Native Simulation Feedback and Tool Access}
\label{sec:tool-ablation}

\paragraph{Ablation Setup.}

To test whether an agent can still complete tasks it already solves once a tool capability is removed, we use Claude Opus~4.8 with Claude Code and select one strong-performing task under Full MCP from each of the ten engineering domains, forming a cross-domain subset. The three tool conditions form two single-factor contrasts: No-Simulation MCP retains MCP model construction, querying, and compilation but disables simulation execution, differing from Full MCP only in simulation feedback; Batch-only removes the MCP interface, differing from Full MCP in the entire MCP toolset.

\paragraph{Results.}

Table~\ref{tab:mcp-ablation} summarizes the three-run mean prerequisite pass rates and overall engineering scores under the three tool conditions. Full MCP versus No-Simulation MCP: artifact delivery, native executability, and engineering qualification fall from 100.00\% to 53.33\%, 33.33\%, and 13.33\%, respectively, and the overall score falls from 78.46 to 7.88; unable to verify its outputs through simulation feedback, the agent scores lower on all ten tasks. Full MCP versus Batch-only: removing MCP lowers the overall score to 18.23, with prerequisite pass rates of 60.00\%, 50.00\%, and 40.00\%, and 8 of 10 tasks score lower.

On these ten tasks that the agent can otherwise complete, both restricted conditions degrade sharply, indicating that simulation feedback and MCP tools are critical for reliably generating executable \simulink{} models. The dimension-level ablation results are reported in Appendix~\ref{app:mcp-ablation-dimensions}.

\begin{table}[!t]
  \centering
  \small
  \caption{\textbf{Ablations of native simulation feedback and tool access.} Results report three-run mean prerequisite pass rates and overall scores for Claude Opus~4.8 with Claude Code on ten cross-domain, high-performing tasks.}
  \label{tab:mcp-ablation}
  \svtablestyle
  \begin{tabular*}{\linewidth}{@{\extracolsep{\fill}}lcccc@{}}
    \toprule
    \svtablehead{Tool condition} & \svtablehead{Artifact delivery} & \svtablehead{Native executability} & \svtablehead{G pass} & \svtablehead{Overall score} \\
    \midrule
    Full MCP & 100.00\% & 100.00\% & 100.00\% & 78.46 \\
    No-Simulation MCP & 53.33\% & 33.33\% & 13.33\% & 7.88 \\
    Batch-only & 60.00\% & 50.00\% & 40.00\% & 18.23 \\
    \bottomrule
  \end{tabular*}
\end{table}

\section{Conclusion}
\label{sec:conclusion}

\benchmark{} evaluates \simulink{} model generation through native simulation and six engineering dimensions across 101 tasks in ten domains. Among six agent systems, the best scores only 42.86, with bottlenecks in both engineering qualification and subsequent performance. Structural similarity does not reliably reflect engineering quality; models closely aligned with the reference can still fail to satisfy engineering requirements, and high-scoring models may still have disordered layouts. Tool ablations on ten selected tasks show substantial benefits from simulation feedback and MCP access. \benchmark{} provides a systematic basis for assessing agents' engineering capabilities and diagnosing failures in executable model generation. The benchmark relies on MATLAB/\simulink{}, and its native simulation-based evaluation can be slow; future work will optimize the evaluation pipeline for greater efficiency and scalability.

\subsection*{AI Use Statement}
Generative AI tools were used in two supporting roles. First, they assisted literature discovery, including identifying related work, by helping formulate and refine search queries, locate candidate papers, and organize the literature around executable engineering benchmarks, \simulink{} model generation, native tool use, and evaluator validation. The authors independently checked the cited works and determined which sources support the manuscript's claims. Second, the tools assisted manuscript drafting and language polishing by proposing wording, translating and condensing author-written passages, and checking consistency in terminology, section and figure labels, and reference formatting. The authors reviewed and verified every AI-assisted output, including all claims, citations, numerical results, and final wording.

\subsection*{Reproducibility Statement}
To support reproducibility, the paper and appendix document task construction, scoring methodology, experimental settings and results, repeated-run protocols, and ablation protocols. The benchmark and results repositories are publicly available at \url{https://github.com/SimuVerity/SimuVerity}. The benchmark repository contains the per-task specifications and executable-system profiles for all 101 tasks, the official simulation scenarios, the evaluator and scoring code, and the reference assets or import recipes; the results repository contains the per-run score tables of all six agent systems and the tool ablation, with their aggregation and validation scripts.

\begingroup
\raggedright
\bibliography{references}
\bibliographystyle{iclr2027_conference}
\endgroup

\clearpage
\appendix
\section*{\centering Appendix}
\begingroup
\renewcommand{\svtablestyle}{\renewcommand{\arraystretch}{1.00}}
\setlength{\intextsep}{8pt}
\setlength{\LTpre}{8pt}
\setlength{\LTpost}{8pt}
\section{Closest-Work Comparison}
\label{app:closest-work}
Table~\ref{tab:closest-work} summarizes the scope and evaluation criteria of the most relevant studies.
\begin{table}[H]
\centering\small
\caption{\textbf{Comparison with related executable-engineering and \simulink{} benchmarks.}}
\label{tab:closest-work}
\svtablestyle
\begin{tabular*}{\linewidth}{@{\extracolsep{\fill}}P{0.18\linewidth}P{0.22\linewidth}P{0.22\linewidth}P{0.28\linewidth}@{}}
\toprule
\svtablehead{Work} & \svtablehead{Task scope} & \svtablehead{Native execution} & \svtablehead{Evaluation focus} \\
\midrule
SLGPT~\citep{shrestha2021slgpt} & \simulink{} file generation & Toolchain execution & Fault discovery \\
SimuGen~\citep{ren2025simugen} & Image-to-model reconstruction & Basic execution & Blocks and connections \\
ASWE-Bench~\citep{abdalla2026aswe} & Automotive control logic & \simulink{} simulation & Boundary and time-series tests \\
SimuBench~\citep{liang2026simuagent} & Multi-domain text-to-model & Basic execution & Reference similarity and completeness \\
VerilogEval~\citep{liu2023verilogeval} & RTL code generation & Transient simulation & Functional correctness \\
RTLLM~\citep{lu2024rtllm} & RTL design generation & Compilation and simulation & Syntax, function, and design quality \\
ModiGen~\citep{xiang2025modigen} & Modelica code generation & Modelica simulation & Simulation success and model quality \\
\svtablehead{\benchmark{}} & Multi-domain executable models & Loading, compilation, simulation & Requirement-grounded A/Q/M/C/R/D evaluation \\
\bottomrule
\end{tabular*}
\end{table}

\section{Task Examples}
\label{app:task-catalog}

\begin{table}[H]
\centering\small
\svtablestyle
\begin{tabular*}{\linewidth}{@{\extracolsep{\fill}}B{0.20\linewidth}P{0.76\linewidth}@{}}
\toprule
\multicolumn{2}{@{}l@{}}{\textbf{XAERO-04: Closed-Loop Tiltrotor VTOL Flight}} \\
\midrule
Engineering goal & Build a VTOL model coupling flight missions, closed-loop control, tilt actuation, and six-degree-of-freedom vehicle dynamics. Support hover waypoint flight, hover-to-fixed-wing transition, and fixed-wing waypoint or orbit flight. \\
Inputs and outputs & Inputs are mission type, initial flight mode, mission profile, payload, wind enablement, and simulation duration. Outputs are position-tracking, transition-state, vehicle-state, and actuator-command time series. \\
Core constraints & Rotor propulsion, independent front-rotor tilt, aerodynamic control surfaces, and vehicle dynamics must form connected paths, with vehicle states participating in feedback control. The model must cover the specified payload and wind conditions, maintain bounded tracking errors and continuous transitions, and avoid prescribed or replayed flight trajectories. \\
Validation scenarios & Six scenarios: nominal hover, payload boundary, transition with wind, forward transition, combined payload and wind, and a fixed-wing turn. \\
Applicable dimensions & A/Q/M/R/D. \\
\bottomrule
\end{tabular*}
\end{table}

The corresponding task-level execution and scoring walkthrough is shown in Figure~\ref{fig:task-walkthrough}.

\begin{table}[H]
\centering\small
\svtablestyle
\begin{tabular*}{\linewidth}{@{\extracolsep{\fill}}B{0.20\linewidth}P{0.76\linewidth}@{}}
\toprule
\multicolumn{2}{@{}l@{}}{\textbf{XAERO-07: Electric-Aircraft Mission and Powertrain}} \\
\midrule
Engineering goal & Build an aircraft powertrain coupling the flight mission, aerodynamic loads, electric propulsion, and power distribution. Evaluate how battery capacity and payload affect mission execution. \\
Inputs and outputs & Inputs are battery capacity and payload. Outputs are flight range, battery state, motor current, and low-voltage auxiliary-load power. \\
Core constraints & Battery capacity affects both stored energy and mass, while payload affects power demand. Electrical and mechanical paths must actively participate in execution. \\
Validation scenarios & Five scenarios: nominal design, increased capacity, increased payload, joint variation, and an intermediate design point. \\
Applicable dimensions & A/Q/M/C/R/D. \\
\bottomrule
\end{tabular*}
\end{table}

\begin{table}[H]
\centering\small
\svtablestyle
\begin{tabular*}{\linewidth}{@{\extracolsep{\fill}}B{0.20\linewidth}P{0.76\linewidth}@{}}
\toprule
\multicolumn{2}{@{}p{\linewidth}@{}}{\textbf{XAUTO-07: Electric Axle Drive with Regenerative Braking}} \\
\midrule
Engineering goal & Build a battery-powered electric axle drive coupling closed-loop motor control, the drivetrain, and longitudinal vehicle dynamics for acceleration and regenerative braking. \\
Inputs and outputs & Inputs scale acceleration and braking demand. Outputs include actual pedal inputs; reference, measured, and estimated torque and torque limits; three-phase currents; and vehicle speed. \\
Core constraints & A high-voltage source, DC bus, controlled three-phase converter, interior permanent-magnet synchronous motor, reduction gear, and vehicle load form a connected electromechanical path. Phase currents and rotor states participate in feedback control. Vehicle speed follows finite-mass vehicle dynamics, without prescribed speed or replayed torque and current trajectories. \\
Validation scenarios & Five scenarios: nominal startup and regeneration transition, reduced acceleration demand, reduced regenerative braking, moderate-demand cyclic operation, and asymmetric pedal excitation. \\
Applicable dimensions & A/Q/M/C/R/D. \\
\bottomrule
\end{tabular*}
\end{table}

\section{Native Scenarios and Scoring Details}
\label{app:evaluator-details}

\subsection{Scenario Coverage}
Each of the 595 formal scenarios is assigned one primary category according to its evaluation focus. Cross-category contributions are not counted twice.

\begin{table}[H]
\setlength{\abovecaptionskip}{0pt}
\caption{\textbf{Formal scenarios across the 101 tasks, counted by primary category.}}
\label{tab:appendix-scenario-coverage}
\centering\small
\svtablestyle
\begin{tabular*}{\linewidth}{@{\extracolsep{\fill}}lr@{}}
\toprule
\svtablehead{Primary scenario category} & \svtablehead{Count} \\
\midrule
Operating-envelope scenarios & 205 \\
Interaction and fault scenarios & 189 \\
Temporal-process scenarios & 79 \\
Mechanism and causal-check scenarios & 122 \\
\midrule
\svtablehead{Total} & 595 \\
\bottomrule
\end{tabular*}
\end{table}

\subsection{Dimension Applicability}
Domain experts determine applicable dimensions from each task's engineering requirements and system characteristics. The same dimension set applies to every agent system evaluated on that task.

\begin{table}[H]
\setlength{\abovecaptionskip}{0pt}
\caption{\textbf{Task counts for the six applicable performance dimensions.}}
\label{tab:appendix-dimension-applicability}
\centering\small
\svtablestyle
\begin{tabular*}{\linewidth}{@{\extracolsep{\fill}}lrrrrrr@{}}
\toprule
\svtablehead{Performance dimension} & \svtablehead{A} & \svtablehead{Q} & \svtablehead{M} & \svtablehead{C} & \svtablehead{R} & \svtablehead{D} \\
\midrule
Applicable tasks & 101 & 101 & 93 & 35 & 101 & 98 \\
\bottomrule
\end{tabular*}
\end{table}

\paragraph{Engineering-qualification gate.}
The G gate evaluates only candidates that have been delivered and execute natively, and returns pass or fail. Its checks are designed by human experts based on the task specification and the executable-system profile, and in general include the presence of the required functional roles, the connectivity of the required signal paths, and the production of valid evaluation evidence according to the task's output contract.

\subsection{Electric-Axle Scoring Example}
The five scenarios of XAUTO-07 cover all four categories. Table~\ref{tab:appendix-xauto-scenarios} lists their main performance associations.

\begin{table}[H]
\setlength{\abovecaptionskip}{0pt}
\caption{\textbf{Electric-axle scenarios, engineering checks, and main performance associations. All scenarios undergo output-completeness checks and contribute to dimension scores and cross-scenario evaluation.}}
\label{tab:appendix-xauto-scenarios}
\centering\small
\setlength{\tabcolsep}{4pt}
\svtablestyle
\begin{tabular*}{\linewidth}{@{\extracolsep{\fill}}P{0.14\linewidth}P{0.23\linewidth}P{0.49\linewidth}P{0.07\linewidth}@{}}
\toprule
\svtablehead{Primary category} & \svtablehead{Scenario} & \svtablehead{Engineering checks} & \svtablehead{Main dims.} \\
\midrule
Temporal process & Nominal startup and regeneration transition & Startup torque establishment, transition timing from motoring to regeneration, and torque tracking. & D \\
Mechanism and causal check & Reduced acceleration demand & Paired comparison with the nominal scenario of torque and speed responses to acceleration demand, together with electromechanical coupling. & M/C \\
Mechanism and causal check & Reduced regenerative braking & Paired comparison with the nominal scenario of generating torque, final vehicle speed, and braking-response delay. & C/D \\
Operating envelope & Moderate-demand cyclic operation & Torque tracking, ripple, and current balance at a moderate operating point within the input range. & A/Q \\
Interaction and fault & Asymmetric pedal excitation & Output quality and performance retention relative to nominal operation under asymmetric acceleration and braking demands. & R \\
\bottomrule
\end{tabular*}
\end{table}

\Needspace{12\baselineskip}
Table~\ref{tab:appendix-xauto-dimensions} reports the six dimension scores and their derivation for the Opus Run~1 candidate on this task.

\begin{table}[H]
\setlength{\abovecaptionskip}{0pt}
\caption{\textbf{Six-dimensional performance and score derivation for the Opus Run~1 electric-axle candidate. Scores are rounded to two decimal places.}}
\label{tab:appendix-xauto-dimensions}
\centering\small
\setlength{\tabcolsep}{4pt}
\svtablestyle
\begin{tabular*}{\linewidth}{@{\extracolsep{\fill}}P{0.05\linewidth}P{0.24\linewidth}P{0.56\linewidth}r@{}}
\toprule
\svtablehead{Dim.} & \svtablehead{Engineering metrics} & \svtablehead{Score derivation for this candidate} & \svtablehead{Score} \\
\midrule
A & Torque-tracking and torque-estimation errors & Measured-to-reference tracking error and estimated-to-measured torque error are converted to utilities, then averaged across the five scenarios. & 100.00 \\
Q & Torque ripple, peak-to-average ratio, phase-current crest factor and balance, and ride smoothness & Torque and current fluctuations, three-phase current imbalance, and vehicle jerk are evaluated. Their utilities are averaged across the five scenarios. & 100.00 \\
M & Functional-path connectivity, motoring and generating operation, and electromechanical coupling & Connectivity, bidirectional operation, torque--acceleration relationships, and current--torque relationships are combined, with greater weight on weaker components. Electromechanical coupling is the main shortfall here. & 78.25 \\
C & Acceleration/braking input fidelity and corresponding vehicle responses & Checks assess whether acceleration and braking inputs are applied as required and whether torque and speed respond accordingly. Weaker input responses receive greater weight. Braking-input fidelity and braking response are weaker here. & 71.88 \\
R & Retention of accuracy, output quality, electromechanical coupling, and bidirectional operation & Performance retention relative to nominal operation is calculated for each scenario, then aggregated through a weighted combination of the minimum and mean. Stronger motoring with light braking gives the lowest retention, although the overall decline is small. & 99.90 \\
D & Startup torque-establishment and regenerative-braking delays & Three delays are checked: nominal startup, nominal braking, and entry into generating operation under reduced braking demand. & 100.00 \\
\bottomrule
\end{tabular*}
\end{table}

After the prerequisite checks, we use a weighted geometric mean so that all applicable dimensions influence the overall score and a high score on one dimension cannot fully compensate for a critical weakness in another. Domain experts set the weights separately for each task according to its engineering objectives, key failure risks, and dimension importance; non-applicable dimensions are excluded from aggregation. For XAUTO-07, the expert-designed weights are $0.25$, $0.10$, $0.25$, $0.20$, $0.10$, and $0.10$ for A/Q/M/C/R/D, respectively, and the task score follows the same form as the main text:
\begin{equation}
S_t(x)=\Gamma_t\!\left(r_t^{\mathrm{del}},r_t^{\mathrm{exe}},r_t^G\right)H_t(\mathbf{v}_t),
\qquad
H_t(\mathbf{v}_t)=A^{0.25}Q^{0.10}M^{0.25}C^{0.20}R^{0.10}D^{0.10}.
\label{eq:appendix-xauto-score}
\end{equation}
This candidate passes all three prerequisite checks, so $\Gamma_t=1$, and obtains 88.03 when its unrounded dimension scores are aggregated.

\subsection{Expert Validation of Evaluator Validity}
\label{subsec:expert-validation}

The sample was stratified into 8 high-score candidates, 5 medium-score candidates, 5 low-but-nonzero candidates, 5 candidates that passed the qualification gate but received an overall score of zero, and 7 candidates that failed the qualification gate. For each candidate model, the experts reviewed the task specification, delivered model, and native simulation outputs, and used a five-point scale to rate overall engineering adequacy: 1 point indicates that the task requirements are clearly unmet; 2 points indicate that most requirements are unmet; 3 points indicate a borderline case or insufficient evidence for a determination; 4 points indicate that the task requirements are largely met; and 5 points indicate that the requirements are clearly met with a complete implementation.

The experts did not see evaluator scores, discuss their ratings, or participate in evaluator design. Across the 30 candidate models, the experts assigned the same score in 21 cases, differed by 1 point in 7 cases, and differed by 2 points in 2 cases, with a mean absolute difference of 0.37 points. The weighted Cohen's $\kappa=0.90$ with a 95\% confidence interval of $[0.79,0.96]$; rank correlations for each comparison are reported in Table~\ref{tab:expert-validation}.

\begin{table}[H]
\setlength{\abovecaptionskip}{0pt}
\caption{\textbf{Rank correlations between evaluator scores and domain-expert ratings.}}
\label{tab:expert-validation}
\centering\small
\setlength{\tabcolsep}{3pt}
\svtablestyle
\begin{tabular*}{\linewidth}{@{\extracolsep{\fill}}P{0.18\linewidth}P{0.47\linewidth}rcr@{}}
\toprule
\svtablehead{Statistic} & \svtablehead{Comparison} & \svtablehead{$n$} & \svtablehead{Estimate} & \svtablehead{95\% CI} \\
\midrule
Spearman $\rho$ & Expert 1 vs. Expert 2 & 30 & 0.87 & [0.71, 0.96] \\
Spearman $\rho$ & Evaluator vs. Expert 1 & 30 & 0.88 & [0.71, 0.96] \\
Spearman $\rho$ & Evaluator vs. Expert 2 & 30 & 0.94 & [0.85, 0.98] \\
Spearman $\rho$ & Evaluator vs. expert mean & 30 & 0.94 & [0.88, 0.98] \\
Spearman $\rho$ & Evaluator vs. expert mean (nonzero candidates) & 18 & 0.88 & [0.72, 0.94] \\
\bottomrule
\end{tabular*}
\end{table}

The table shows high agreement and rank consistency: evaluator scores correlate strongly with both individual experts and their mean rating, and the correlation remains high after excluding zero-score candidates. Thus, evaluator rankings closely agree with expert judgments and are not driven mainly by the zero-score cases.

\section{Experimental Settings and Supplementary Results}
\label{app:full-results}

\subsection{Implementation Settings}
\label{subsec:implementation-settings}

\begin{table}[H]
\setlength{\abovecaptionskip}{0pt}
\caption{\textbf{Reasoning effort and context windows. Each model uses its highest supported reasoning level and its official native context-window setting.}}
\label{tab:appendix-settings}
\centering\small
\setlength{\tabcolsep}{3pt}
\svtablestyle
\begin{tabular*}{\linewidth}{@{\extracolsep{\fill}}lcccccc@{}}
\toprule
\svtablehead{Setting} & \svtablehead{\shortstack{Claude\\Opus 4.8}} & \svtablehead{GPT-5.5} & \svtablehead{\shortstack{DeepSeek\\V4-Pro}} & \svtablehead{\shortstack{Qwen3.8\\Max}} & \svtablehead{\shortstack{GLM-5.3\\Flash}} & \svtablehead{\shortstack{Qwen3.8\\27B-FP8}} \\
\midrule
Reasoning effort & \texttt{max} & \texttt{xhigh} & \texttt{max} & \texttt{xhigh} & \texttt{max} & \texttt{xhigh} \\
Context window (tokens) & 1M & 1,050,000 & 1M & 1M & 1,048,576 & 262,144 \\
\bottomrule
\end{tabular*}
\end{table}

The agent frameworks are Claude Code~2.1.238 and Codex~0.149.0. Native execution uses MATLAB/\simulink{} R2025a and the products required by each task.

\paragraph{Harness pairing and stopping rule.}
Every model that can be called through an Anthropic-compatible interface runs in the same harness (Claude Code); GPT-5.5 uses its native Codex. Opus~4.8 and GPT-5.5 therefore both run in their vendors' native agent tools, and the unit of evaluation is the agent system (model + harness). Sampling parameters follow each harness's defaults. Each run stops when the agent declares completion.

\paragraph{Environment isolation.}
Outbound connections of the agent process and the MATLAB session are blocked, and the harness's built-in web tools are disabled. The \texttt{examples} directory under the MATLAB root, the \texttt{examples} directories of the toolboxes, and the support-package directories are removed. Reference \texttt{.slx} files, scenario definitions, and scoring scripts are stored in directories the agent cannot access and are read only by the evaluation process after the agent run ends. We searched all agent trajectories for attempts to access example directories, reference files, or the network; none succeeded.

\subsection{Statistical Reliability}

Following~\citet{miller2024errorbars}, we report uncertainty at the task level. Each task score is the mean over three independent runs, and the system score is the mean over 101 task scores. We obtain 95\% confidence intervals from 10,000 bootstrap resamples of tasks with replacement and report percentile intervals. For pairwise comparisons, we resample paired task-level differences and compute $p$-values using the Wilcoxon signed-rank test. These intervals quantify uncertainty when resampling similar tasks; run-to-run variation is shown in the Run 1--3 columns.

\begin{table}[H]
\setlength{\abovecaptionskip}{0pt}
\caption{\textbf{Scores for the three runs, mean, task-level 95\% confidence interval, and the proportion of tasks with consistent G outcomes across runs.}}
\label{tab:appendix-main-reliability}
\centering\small
\setlength{\tabcolsep}{3pt}
\svtablestyle
\begin{tabular*}{\linewidth}{@{\extracolsep{\fill}}lrrrrcr@{}}
\toprule
\svtablehead{System} & \svtablehead{Run 1} & \svtablehead{Run 2} & \svtablehead{Run 3} & \svtablehead{Mean} & \svtablehead{95\% CI} & \svtablehead{Consistent G} \\
\midrule
Claude Opus 4.8 & 44.34 & 41.05 & 43.18 & 42.86 & [37.03, 48.73] & 92.08\% \\
GPT-5.5 & 39.40 & 42.87 & 42.91 & 41.72 & [35.29, 48.31] & 89.11\% \\
DeepSeek-V4-Pro & 27.35 & 29.52 & 28.32 & 28.40 & [22.01, 34.76] & 82.18\% \\
Qwen3.8-Max & 24.57 & 25.85 & 24.62 & 25.01 & [18.85, 31.37] & 86.14\% \\
GLM-5.3-Flash & 3.36 & 6.19 & 5.38 & 4.98 & [2.40, 8.04] & 81.19\% \\
Qwen3.8-27B-FP8 & 1.58 & 1.56 & 1.67 & 1.60 & [0.41, 2.99] & 89.11\% \\
\bottomrule
\end{tabular*}
\end{table}

Across the three runs, only Claude Opus 4.8 and GPT-5.5 exchange positions. Paired task-level comparisons place the six agent systems into four tiers: $\{\text{Claude Opus 4.8},\ \text{GPT-5.5}\} > \{\text{DeepSeek-V4-Pro},\ \text{Qwen3.8-Max}\} > \text{GLM-5.3-Flash} > \text{Qwen3.8-27B-FP8}$. The within-tier pairwise differences are not significant ($p=0.86$ and $0.46$), whereas all adjacent between-tier differences are significant ($p\le 0.012$). The proportion of tasks with consistent G outcomes across runs is 81.19--92.08\%.

\begin{table}[H]
\setlength{\abovecaptionskip}{0pt}
\caption{\textbf{Differences between adjacent agent systems. 95\% confidence intervals are obtained by paired task-level bootstrap, and $p$-values are from the Wilcoxon signed-rank test.}}
\label{tab:appendix-tier-contrasts}
\centering\small
\svtablestyle
\begin{tabular*}{\linewidth}{@{\extracolsep{\fill}}lrrr@{}}
\toprule
\svtablehead{Comparison} & \svtablehead{Difference} & \svtablehead{95\% CI} & \svtablehead{$p$} \\
\midrule
Claude Opus 4.8 -- GPT-5.5 & 1.13 & [-5.80, 7.68] & 0.86 \\
GPT-5.5 -- DeepSeek-V4-Pro & 13.33 & [6.93, 19.89] & $<0.001$ \\
DeepSeek-V4-Pro -- Qwen3.8-Max & 3.38 & [-3.63, 10.39] & 0.46 \\
Qwen3.8-Max -- GLM-5.3-Flash & 20.04 & [13.73, 26.54] & $<0.001$ \\
GLM-5.3-Flash -- Qwen3.8-27B-FP8 & 3.37 & [0.92, 6.03] & 0.012 \\
\bottomrule
\end{tabular*}
\end{table}

System ranking is insensitive to the aggregation rule. Table~\ref{tab:appendix-aggregation-sensitivity} recomputes task scores with the G gate retained, using the arithmetic mean, equal-weight geometric mean, or minimum of applicable dimension scores. The ordering below the top pair is unchanged, and Kendall's $\tau$ with the original ranking is 0.87, 1.00, and 0.87, respectively.

\begin{table}[H]
\setlength{\abovecaptionskip}{0pt}
\caption{\textbf{Overall scores under the original and three alternative aggregation rules; candidates failing G receive zero.}}
\label{tab:appendix-aggregation-sensitivity}
\centering\small
\svtablestyle
\begin{tabular*}{\linewidth}{@{\extracolsep{\fill}}lrrrr@{}}
\toprule
\svtablehead{System} & \svtablehead{Original} & \svtablehead{Arithmetic} & \svtablehead{Equal-weight geometric} & \svtablehead{Minimum} \\
\midrule
Claude Opus 4.8 & 42.86 & 50.34 & 44.00 & 25.38 \\
GPT-5.5 & 41.72 & 50.65 & 42.49 & 26.48 \\
DeepSeek-V4-Pro & 28.40 & 37.70 & 29.40 & 18.87 \\
Qwen3.8-Max & 25.01 & 33.01 & 25.61 & 16.06 \\
GLM-5.3-Flash & 4.98 & 7.37 & 5.38 & 3.20 \\
Qwen3.8-27B-FP8 & 1.60 & 3.29 & 1.67 & 0.74 \\
\bottomrule
\end{tabular*}
\end{table}

\begin{table}[H]
\setlength{\abovecaptionskip}{0pt}
\caption{\textbf{Scores for the three runs, mean, task-level 95\% confidence interval, and tasks with scores below Full MCP on the tool-ablation subset.}}
\label{tab:appendix-ablation-reliability}
\centering\small
\svtablestyle
\setlength{\tabcolsep}{3pt}
\begin{tabular*}{\linewidth}{@{\extracolsep{\fill}}lrrrrcr@{}}
\toprule
\svtablehead{Condition} & \svtablehead{Run 1} & \svtablehead{Run 2} & \svtablehead{Run 3} & \svtablehead{Mean} & \svtablehead{95\% CI} & \svtablehead{Tasks below Full MCP} \\
\midrule
Full MCP & 82.19 & 76.04 & 77.16 & 78.46 & [76.13, 80.93] & --- \\
No-Simulation MCP & 6.63 & 10.85 & 6.17 & 7.88 & [0.00, 20.06] & 10/10 \\
Batch-only & 16.81 & 20.67 & 17.22 & 18.23 & [0.00, 40.73] & 8/10 \\
\bottomrule
\end{tabular*}
\end{table}

Full MCP exceeds No-Simulation MCP by 70.58 points (95\% CI [58.02, 78.63], Wilcoxon $p=0.002$) and Batch-only by 60.23 points ([37.13, 77.58], $p=0.010$). At the task level, they are lower on 10/10 and 8/10 tasks, respectively.

In summary, the six agent systems fall into four tiers, with significant differences between adjacent tiers. In every run, the mean scores of both restricted conditions are below Full MCP. The conclusion remains consistent across the three runs and continues to hold under task-level bootstrap resampling.

\subsection{Cross-System Domain Results}

Table~\ref{tab:appendix-domain-results} reports end-to-end and post-G overall scores across ten domains for six systems, comparing overall performance and domain performance after qualification.

\begin{table}[H]
\setlength{\abovecaptionskip}{0pt}
\caption{\textbf{End-to-end and post-G overall scores across ten domains for six systems.} An em dash indicates no qualified candidate in any run; 0.00 denotes a defined zero score.}
\label{tab:appendix-domain-results}
\centering\small
\setlength{\tabcolsep}{3pt}
\svtablestyle
\begin{tabular*}{\linewidth}{@{\extracolsep{\fill}}lrrrrrr@{}}
\toprule
\multicolumn{7}{c}{\textbf{(a) End-to-end overall score}} \\
\svtablehead{Domain} & \svtablehead{\shortstack{Claude\\Opus 4.8}} & \svtablehead{GPT-5.5} & \svtablehead{\shortstack{DeepSeek\\V4-Pro}} & \svtablehead{\shortstack{Qwen3.8\\Max}} & \svtablehead{\shortstack{GLM-5.3\\Flash}} & \svtablehead{\shortstack{Qwen3.8\\27B-FP8}} \\
\midrule
Power electronics \& control & 25.99 & 30.51 & 21.63 & 1.79 & 5.72 & 0.00 \\
Mechanical & 32.85 & 34.25 & 23.74 & 35.91 & 0.00 & 0.00 \\
Robotics & 55.49 & 49.79 & 47.75 & 36.83 & 0.00 & 0.00 \\
Thermal & 49.44 & 59.48 & 36.13 & 36.52 & 3.52 & 0.00 \\
Fluid & 48.92 & 46.35 & 39.79 & 24.81 & 19.48 & 5.87 \\
Automotive & 53.03 & 43.95 & 15.19 & 18.29 & 3.64 & 3.04 \\
Battery & 33.13 & 36.94 & 31.04 & 5.16 & 10.47 & 0.00 \\
Aerospace & 42.49 & 35.29 & 6.92 & 21.51 & 2.52 & 0.00 \\
Communication & 33.16 & 27.56 & 35.34 & 29.41 & 3.14 & 3.76 \\
Biomedical & 50.91 & 49.68 & 26.13 & 37.76 & 0.00 & 3.33 \\
\midrule
\multicolumn{7}{c}{\textbf{(b) Post-G overall score}} \\
\svtablehead{Domain} & \svtablehead{\shortstack{Claude\\Opus 4.8}} & \svtablehead{GPT-5.5} & \svtablehead{\shortstack{DeepSeek\\V4-Pro}} & \svtablehead{\shortstack{Qwen3.8\\Max}} & \svtablehead{\shortstack{GLM-5.3\\Flash}} & \svtablehead{\shortstack{Qwen3.8\\27B-FP8}} \\
\midrule
Power electronics \& control & 58.49 & 54.93 & 64.64 & 16.10 & 77.24 & 0.00 \\
Mechanical & 47.46 & 59.24 & 46.18 & 49.38 & --- & --- \\
Robotics & 64.05 & 55.33 & 54.92 & 50.58 & 0.00 & 0.00 \\
Thermal & 62.90 & 69.99 & 56.78 & 57.76 & 26.24 & 0.00 \\
Fluid & 67.26 & 80.50 & 65.83 & 74.88 & 79.92 & 41.50 \\
Automotive & 58.33 & 63.47 & 26.09 & 50.29 & 13.35 & 25.38 \\
Battery & 38.29 & 44.99 & 36.00 & 9.75 & 43.38 & --- \\
Aerospace & 47.21 & 35.29 & 15.60 & 37.73 & 37.87 & --- \\
Communication & 50.08 & 35.43 & 43.40 & 36.14 & 28.24 & 38.90 \\
Biomedical & 65.45 & 61.18 & 58.78 & 68.72 & 0.00 & 29.96 \\
\bottomrule
\end{tabular*}
\end{table}

End-to-end results show that the six systems have different domain-performance profiles. Post-G results further show substantial differences between systems across domains even after engineering qualification.

Claude Opus 4.8's cross-domain capability patterns are analyzed in Section~\ref{sec:cross-domain}; the following paragraph supplements the results for the other agent systems. GPT-5.5 achieves relatively higher end-to-end scores in thermal, robotics, biomedical, and fluid domains, but lower scores in communication and in power electronics and control; after engineering qualification, its stronger domains are fluid and thermal. DeepSeek-V4-Pro is relatively stronger in robotics and fluid, but substantially constrained in aerospace and automotive; its qualified candidates remain comparatively strong in power electronics and control, fluid, and biomedical domains. Qwen3.8-Max's end-to-end scores are concentrated in mechanical, robotics, thermal, and biomedical domains, while power electronics and control and battery remain low; after qualification, fluid and biomedical are stronger, whereas power electronics and control and battery remain constrained. GLM-5.3-Flash and Qwen3.8-27B-FP8 obtain near-zero end-to-end scores in most domains, and several domains lack qualified candidates, indicating that these agent systems have not yet formed stable cross-domain engineering capability.

\begin{table}[H]
  \caption{\textbf{Cross-domain engineering profiles for Claude Opus 4.8 with Claude Code.} ``Post-G'' averages include only candidates that pass engineering qualification; an em dash marks domains with no dominant bottleneck type.}
  \label{tab:domain-regimes}
  \centering
  \small
  \setlength{\tabcolsep}{4.5pt}
  \svtablestyle
  \begin{tabular*}{\linewidth}{@{\extracolsep{\fill}}lrrrrl@{}}
    \toprule
    \svtablehead{Engineering domain} & \svtablehead{$N$} & \svtablehead{Overall} & \svtablehead{G pass} & \svtablehead{Post-G} & \svtablehead{Capability regime} \\
    \midrule
    Power electronics \& control & 9  & 25.99 & 44.44\% & 58.49 & Qualification dominated \\
    Mechanical systems            & 11 & 32.85 & 69.70\% & 47.46 & Compound \\
    Robotics                      & 10 & 55.49 & 86.67\% & 64.05 & --- \\
    Thermal systems               & 11 & 49.44 & 78.79\% & 62.90 & --- \\
    Fluid systems                 & 11 & 48.92 & 72.73\% & 67.26 & Qualification dominated \\
    Automotive systems            & 11 & 53.03 & 90.91\% & 58.33 & --- \\
    Battery systems               & 10 & 33.13 & 86.67\% & 38.29 & Post-G performance \\
    Aerospace systems             & 10 & 42.49 & 90.00\% & 47.21 & Post-G performance \\
    Communication systems         & 9  & 33.16 & 66.67\% & 50.08 & Compound \\
    Biomedical systems            & 9  & 50.91 & 77.78\% & 65.45 & --- \\
    \bottomrule
  \end{tabular*}
\end{table}

\begin{figure}[H]
  \centering
  \includegraphics[width=\linewidth]{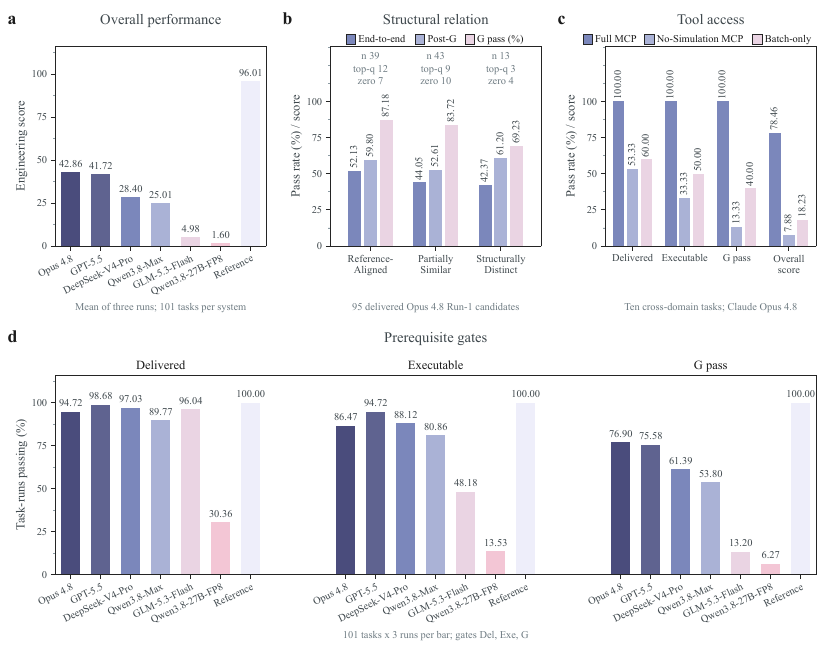}
  \caption{Summary of the experimental results. (a) Overall performance, (b) structural relation, (c) tool access, and (d) prerequisite-gate outcomes.}
  \label{fig:appendix-experimental-summary}
\end{figure}

\section{Structural Relations and Representative Cases}
\label{app:evaluator-controls}

\subsection{Six-Dimensional Performance}

\begin{table}[H]
\setlength{\abovecaptionskip}{0pt}
\caption{\textbf{Six-dimensional scores of the four representative candidates.} All four candidates pass the G gate; an em dash indicates a non-applicable dimension.}
\label{tab:appendix-structural-dimensions}
\centering\small
\setlength{\tabcolsep}{4pt}
\svtablestyle
\begin{tabular*}{\linewidth}{@{\extracolsep{\fill}}lrrrrrr@{}}
\toprule
\svtablehead{Case} & \svtablehead{A} & \svtablehead{Q} & \svtablehead{M} & \svtablehead{C} & \svtablehead{R} & \svtablehead{D} \\
\midrule
Wheel-loader coordination (\path{XMECH-11}) & 95.55 & 99.89 & 100.00 & 50.00 & 91.40 & 100.00 \\
Electric-aircraft mission and powertrain (\path{XAERO-07}) & 100.00 & 33.33 & 75.00 & 100.00 & 66.67 & 100.00 \\
Serial robotic arm (\path{XMECH-10}) & 0.00 & 0.00 & 0.00 & --- & 0.00 & 0.00 \\
Radar system (\path{XCOMM-01}) & 0.00 & 55.18 & 25.00 & --- & 0.00 & 63.69 \\
\bottomrule
\end{tabular*}
\end{table}

\subsection{Structural Features and Engineering Outcomes}

Structural relations are assigned qualitatively from block choices and abstraction level, system topology, subsystem decomposition, and signal and feedback routing. The three mutually exclusive relations are defined below.

\begin{table}[H]
  \centering
  \small
  \caption{\textbf{Three-level structural relation criteria.}}
  \label{tab:structural-relation-criteria}
  \setlength{\tabcolsep}{4pt}
  \svtablestyle
  \begin{tabular*}{\linewidth}{@{\extracolsep{\fill}}P{0.24\linewidth}P{0.70\linewidth}@{}}
    \toprule
    \svtablehead{Relation} & \svtablehead{Decision rule} \\
    \midrule
    Reference-Aligned (RA) & The candidate preserves a recognizable correspondence in major block choices and abstraction level, system topology, subsystem decomposition, and signal/feedback routing. Differences are limited mainly to local substitutions, parameterization, auxiliary blocks, or bounded reorganization. \\
    Partially Similar (PS) & The candidate retains some central components or engineering paths, but materially changes the overall architecture, abstraction level, subsystem organization, or signal/feedback routing. Similarity is local rather than system-wide. \\
    Structurally Distinct (SD) & The candidate differs substantially in system-level formulation, abstraction level, topology, subsystem organization, or signal/feedback routing. Shared elements are mainly task-mandated interfaces, generic domain components, or the engineering objective. \\
    \bottomrule
  \end{tabular*}
  \vspace{2pt}
  \parbox{\linewidth}{\footnotesize Human domain experts classify candidate results from Opus 4.8 Run~1 according to the qualitative criteria above. Layout coordinates, colors, labels, parameter values, solver settings, and block-count differences alone do not determine a label; the relation is not part of the \benchmark{} score.}
\end{table}

The following table gives representative valid-diversity and deceptive-similarity cases from the structural-similarity analysis in Section~\ref{sec:evaluator-results}.

\begin{table}[H]
  \centering
  \small
  \caption{\textbf{Representative cases illustrating valid structural diversity and deceptive structural similarity.}}
  \label{tab:similarity-cases}
  \setlength{\tabcolsep}{4pt}
  \svtablestyle
  \begin{tabular*}{\linewidth}{@{\extracolsep{\fill}}>{\raggedright\arraybackslash}p{0.13\linewidth}>{\raggedright\arraybackslash}p{0.19\linewidth}>{\raggedright\arraybackslash}p{0.13\linewidth}cr>{\raggedright\arraybackslash}p{0.34\linewidth}@{}}
    \toprule
    \svtablehead{Phenomenon} & \svtablehead{Task} & \svtablehead{Relation} & \svtablehead{G} & \svtablehead{Score} & \svtablehead{Engineering evidence} \\
    \midrule
    Valid diversity & \path{XMECH-11} & Structurally Distinct & Pass & 85.10 & Reference: detailed physical loader, engine, transmission, and surface subsystems; candidate: a lumped MATLAB-function plant, a different system-level structural implementation. \\
    Valid diversity & \path{XAERO-07} & Structurally Distinct & Pass & 80.07 & Reference: physical Simscape aircraft distribution network; candidate: signal-level lumped equations, with different component families, topology, and routing. \\
    Deceptive similarity & \path{XMECH-10} & Reference-Aligned & Pass & 0.00 & A recognizable serial youBot arm topology extends from the base pivot to the two-finger gripper, but defects in its hierarchy, actuation, and routing cause severely inadequate performance; all performance dimensions are zero. \\
    Deceptive similarity & \path{XCOMM-01} & Reference-Aligned & Pass & 0.00 & The organization and output paths resemble a single-core radar, but target-range and velocity estimation are defective. Objective accuracy and operating-domain robustness are zero. \\
    \bottomrule
  \end{tabular*}
\end{table}

\section{Additional Visual Layout Interventions}
\label{app:visual-layout}

This appendix provides the seven layout-only comparisons not shown in Figure~\ref{fig:visual-layout}. Within each pair, the block set, parameters, ports, connection topology, hierarchy, simulation configuration, and solver are held fixed; only top-level block and subsystem positions and the routing of existing lines are changed.

\subsection{Supplementary Metric Conventions}
\label{app:layout-metrics}

\begin{itemize}
  \item \textbf{Block overlap:} We count positive-area overlaps among top-level blocks, count each block once, and exclude boundary contact alone.
  \item \textbf{Routing detour:} We count only traceable directed signal paths, excluding paths with zero endpoint distance and undirected physical connections.
  \item \textbf{Line crossings:} We count only internal crossings between different line networks outside blocks, excluding shared endpoints, branches within the same network, and collinear overlap.
\end{itemize}

\subsection{Layout-Only Intervention Results}

For each severely disordered model, experts produced a reorganized version by modifying only top-level block positions, subsystem positions, and existing line routes. Blocks, parameters, ports, connectivity, hierarchy, simulation settings, and solver configuration were kept unchanged.

\begin{table}[H]
  \caption{\textbf{Layout-only optimization of eight high-performing but severely disordered \simulink{} models. Lower values are better for all three layout indicators.}}
  \label{tab:visual-layout}
  \centering
  \footnotesize
  \setlength{\tabcolsep}{2.2pt}
  \svtablestyle
  \begin{tabular*}{\linewidth}{@{\extracolsep{\fill}}>{\raggedright\arraybackslash}p{0.27\linewidth}>{\raggedright\arraybackslash}p{0.15\linewidth}rccc@{}}
    \toprule
    \svtablehead{Candidate model} & \svtablehead{Domain} & \svtablehead{Score} & \svtablehead{Overlap} & \svtablehead{Detour} & \svtablehead{\shortstack{Crossings /\\100 segments}} \\
    \midrule
    QPSK Communication Link & Communication & 83.33 & 0.950$\rightarrow$0.000 & 1.283$\rightarrow$1.009 & 71.0$\rightarrow$19.6 \\
    Two-Body Point-Absorber Wave-Energy Converter & Fluid & 84.22 & 0.966$\rightarrow$0.000 & 5.998$\rightarrow$1.060 & 295.6$\rightarrow$70.8 \\
    Closed-Loop CPU Cooling System & Thermal & 81.02 & 0.907$\rightarrow$0.000 & 1.493$\rightarrow$1.078 & 204.8$\rightarrow$30.6 \\
    Grid-Connected Two-Level Inverter & Power electronics \& control & 74.81 & 0.240$\rightarrow$0.000 & 1.085$\rightarrow$1.080 & 55.9$\rightarrow$43.2 \\
    Dual-Patient Ventilation System & Biomedical & 73.46 & 0.952$\rightarrow$0.000 & 1.388$\rightarrow$1.013 & 120.1$\rightarrow$36.7 \\
    Power-Split Hybrid-Electric Vehicle & Automotive & 72.48 & 0.918$\rightarrow$0.000 & 1.040$\rightarrow$1.011 & 309.9$\rightarrow$70.1 \\
    Robotic Ball-Balancing System & Robotics & 70.89 & 0.324$\rightarrow$0.000 & 1.000$\rightarrow$1.035 & 82.4$\rightarrow$23.2 \\
    Three-Level NPC Inverter & Power electronics \& control & 70.40 & 0.971$\rightarrow$0.000 & 4.226$\rightarrow$1.100 & 142.3$\rightarrow$41.9 \\
    \bottomrule
  \end{tabular*}
\end{table}
\Needspace{20\baselineskip}
\subsection{Seven Supplementary Figures}

\begingroup
\setlength{\intextsep}{6pt}

\begin{figure}[H]
  \centering
  \includegraphics[width=\textwidth]{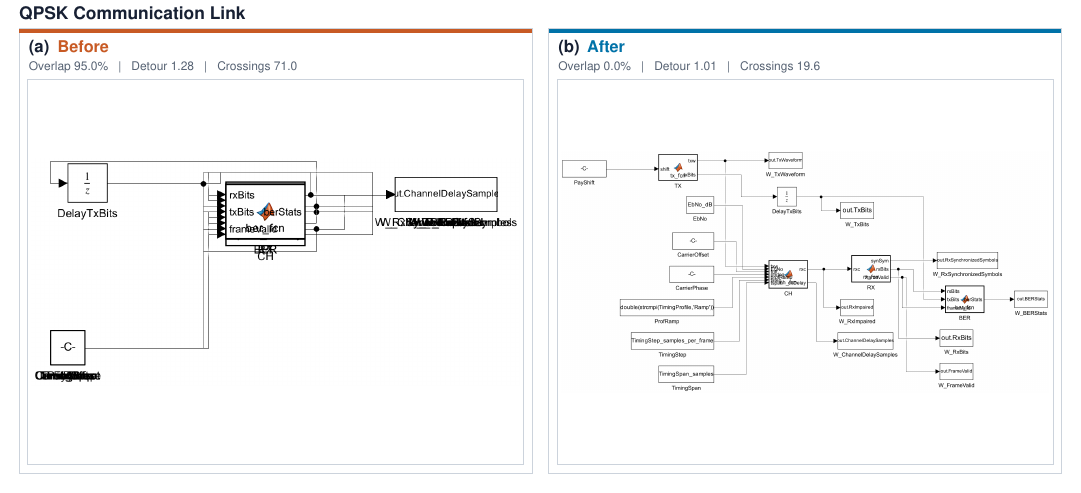}
  \caption{QPSK Communication Link}
  \label{fig:visual-layout-qpsk}
\end{figure}

\begin{figure}[H]
  \centering
  \includegraphics[width=\textwidth]{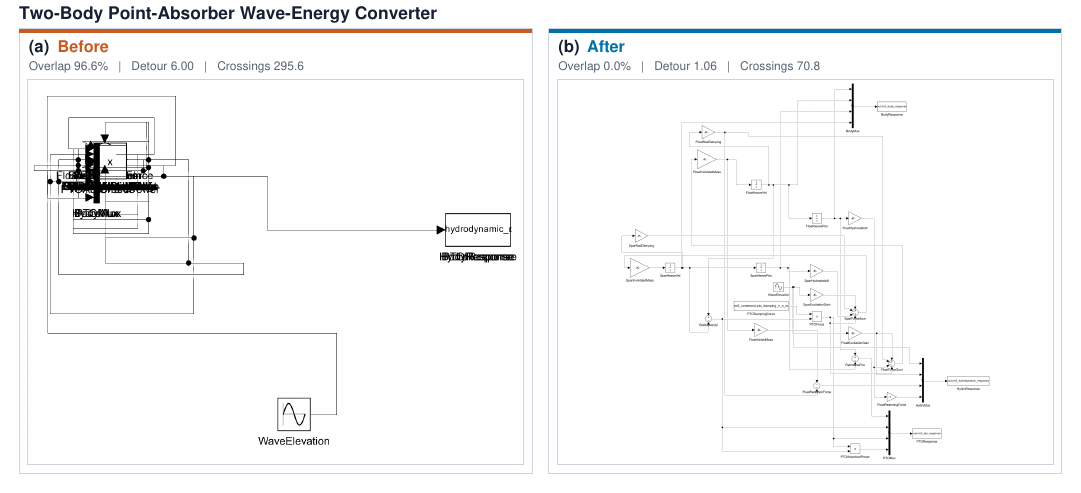}
  \caption{Two-Body Point-Absorber Wave-Energy Converter}
  \label{fig:visual-layout-wec}
\end{figure}

\begin{figure}[H]
  \centering
  \includegraphics[width=\textwidth]{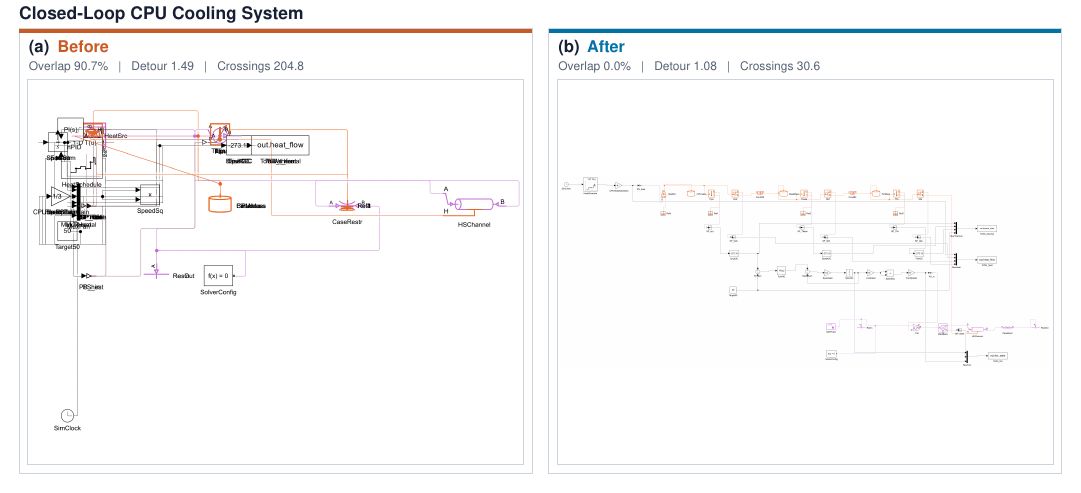}
  \caption{Closed-Loop CPU Cooling System}
  \label{fig:visual-layout-cpu-cooling}
\end{figure}

\begin{figure}[H]
  \centering
  \includegraphics[width=\textwidth]{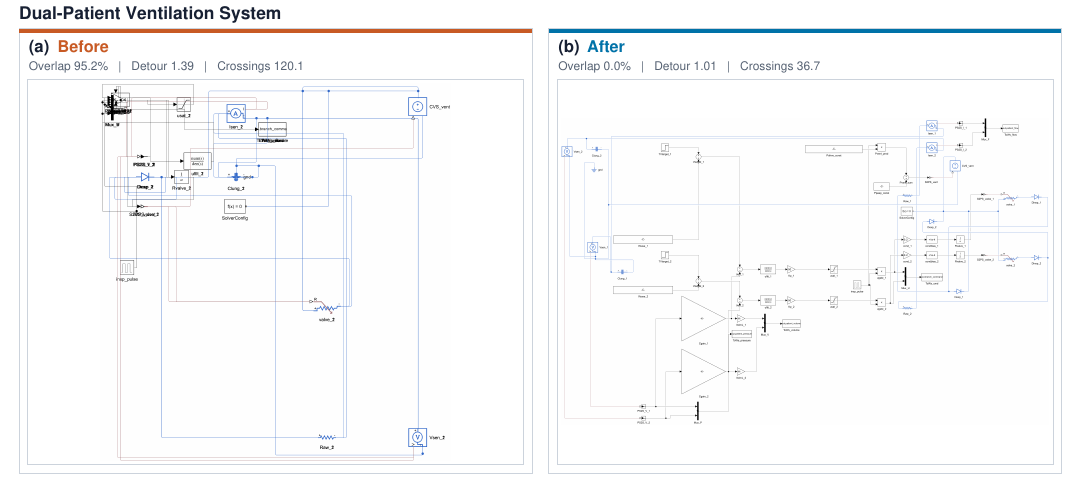}
  \caption{Dual-Patient Ventilation System}
  \label{fig:visual-layout-ventilation}
\end{figure}

\begin{figure}[H]
  \centering
  \includegraphics[width=\textwidth]{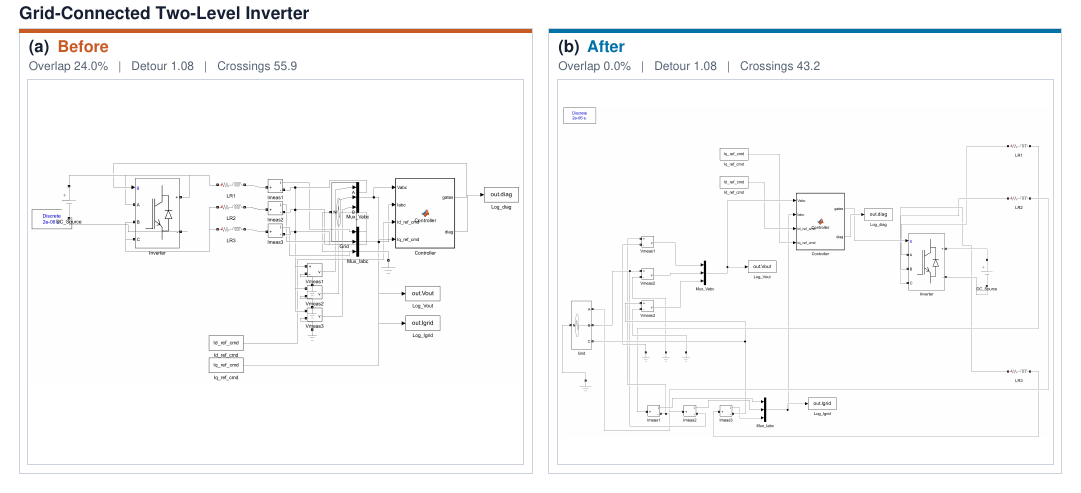}
  \caption{Grid-Connected Two-Level Inverter}
  \label{fig:visual-layout-two-level-inverter}
\end{figure}

\begin{figure}[H]
  \centering
  \includegraphics[width=\textwidth]{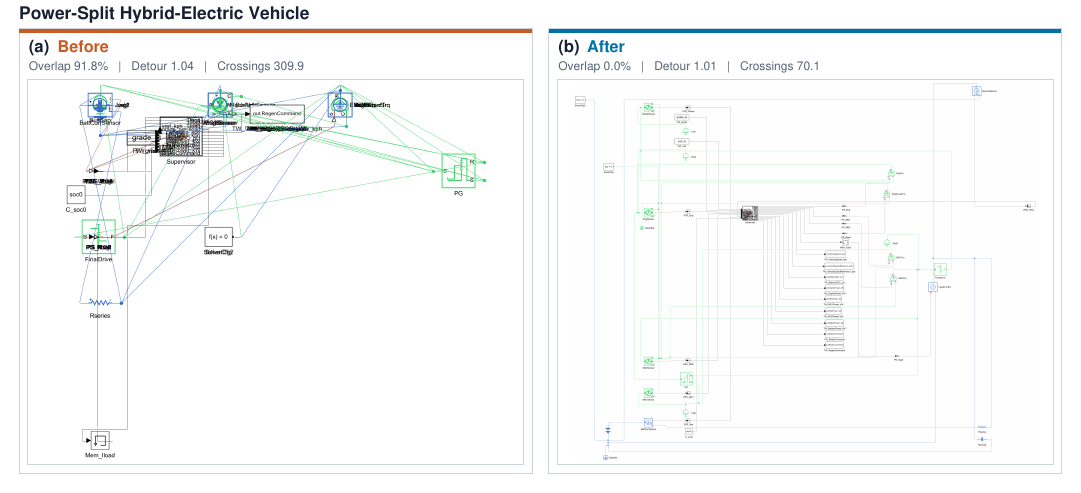}
  \caption{Power-Split Hybrid-Electric Vehicle}
  \label{fig:visual-layout-hybrid-vehicle}
\end{figure}

\begin{figure}[H]
  \centering
  \includegraphics[width=\textwidth]{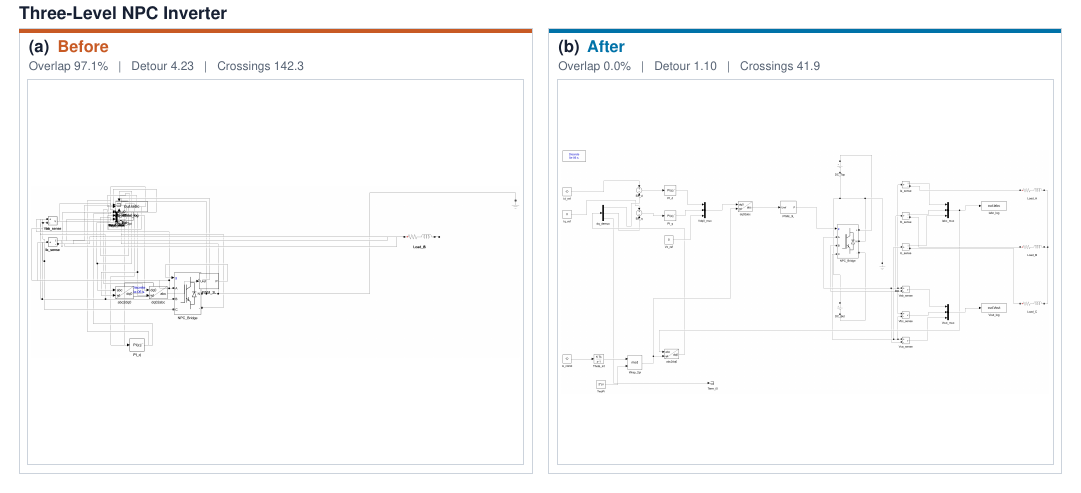}
  \caption{Three-Level NPC Inverter}
  \label{fig:visual-layout-npc-inverter}
\end{figure}
\endgroup

\section{Tool-Ablation Details}
\label{app:mcp-ablation-dimensions}

\subsection{Fixed Tasks and Tool Conditions}

\textbf{Full MCP:} Reuses the three-run mean results of these ten tasks for Opus~4.8 from the main experiment.

\textbf{No-Simulation MCP:} Retains MATLAB MCP capabilities for model creation, reading, editing, querying, structural checks, updating, and compilation, but disables agent-side simulation self-checks during generation: the MCP server rejects simulation calls, including \texttt{sim} invocations issued through the code-execution tool.

\textbf{Batch-only:} Removes access to all MATLAB MCP tools, retains standard file and shell tools, and permits stateless MATLAB batch scripts through shell calls.

\begin{table}[H]
  \centering
  \small
  \caption{\textbf{Task-level engineering scores under three tool conditions across ten cross-domain tasks. Each entry is the arithmetic mean over three runs.}}
  \label{tab:mcp-ablation-task-results}
  \svtablestyle
  \begin{tabular*}{\linewidth}{@{\extracolsep{\fill}}llrrr@{}}
    \toprule
    \svtablehead{Task} & \svtablehead{Domain} & \svtablehead{Full MCP} & \svtablehead{No-Simulation MCP} & \svtablehead{Batch-only} \\
    \midrule
    XELEC-07 & Power electronics \& control & 74.55 & 0.00 & 0.00 \\
    XAERO-07 & Aerospace systems & 75.35 & 0.00 & 0.00 \\
    XAUTO-07 & Automotive systems & 85.13 & 0.00 & 21.49 \\
    XBATT-05 & Battery systems & 73.33 & 0.00 & 75.01 \\
    XCOMM-06 & Communication systems & 79.91 & 60.87 & 0.00 \\
    XFLUID-08 & Fluid systems & 79.78 & 0.00 & 85.81 \\
    XMECH-06 & Mechanical systems & 79.75 & 0.00 & 0.00 \\
    XMED-10 & Biomedical systems & 78.97 & 0.00 & 0.00 \\
    XROBO-02 & Robotics & 83.70 & 17.97 & 0.00 \\
    XTHERM-06 & Thermal systems & 74.14 & 0.00 & 0.00 \\
    \bottomrule
  \end{tabular*}
\end{table}

\subsection{Supplementary Six-Dimension Performance}

\begin{table}[H]
  \centering
  \small
  \caption{\textbf{End-to-end six-dimensional tool-ablation results for Claude Opus~4.8 with Claude Code on ten cross-domain tasks.}}
  \label{tab:mcp-ablation-dimensions}
  \svtablestyle
  \begin{tabular*}{\linewidth}{@{\extracolsep{\fill}}lrrrrrr@{}}
    \toprule
    \svtablehead{Tool condition} & \svtablehead{A} & \svtablehead{Q} & \svtablehead{M} & \svtablehead{C} & \svtablehead{R} & \svtablehead{D} \\
    \midrule
    Full MCP & 78.09 & 86.82 & 81.35 & 97.08 & 84.50 & 75.58 \\
    No-Simulation MCP & 6.56 & 9.71 & 14.07 & 0.00 & 8.38 & 5.86 \\
    Batch-only & 21.70 & 32.89 & 38.53 & 18.76 & 26.08 & 21.12 \\
    \bottomrule
  \end{tabular*}
\end{table}

\endgroup

\end{document}